\documentclass[aps,preprint,prd,showpacs,nofootinbib]{revtex4}
\usepackage{amsmath}
\usepackage{graphicx}
\usepackage{dcolumn}
\usepackage{bm}
\usepackage{amssymb}
\usepackage{latexsym}

\def\be{\begin{equation}}
\def\ee{\end{equation}}
\def\ba{\begin{eqnarray}}
\def\ea{\end{eqnarray}}

\begin{document}

\title{Inflation in Oscillating Universe   }

\author{Yun-Song Piao$^{a,b}$}
\author{Yuan-Zhong Zhang$^{c,d}$}
\affiliation{${}^a$Institute of High Energy Physics, Chinese
Academy of Science, P.O. Box 918-4, Beijing 100039, P. R. China}
\affiliation{${}^b$Interdisciplinary Center of Theoretical
Studies, Chinese Academy of Sciences, P.O. Box 2735, Beijing
100080, China} \affiliation{${}^c$CCAST (World Lab.), P.O. Box
8730, Beijing 100080} \affiliation{${}^d$Institute of Theoretical
Physics, Chinese Academy of Sciences, P.O. Box 2735, Beijing
100080, China }

\begin{abstract}

We make use of possible high energy correction to the Friedmann
equation to implement the bounce and study the behavior of massive
scalar field before and after bounce semianalytically and
numerically. We find that the slow-roll inflation can be preceded
by the kinetic dominated contraction. During this process, the
field initially in the bottom of its potential can be driven by
the anti-frictional force resulted from the contraction and roll
up its potential hill, and when it rolls down after the bounce, it
can driven a period inflation. The required e-folds number during
the inflation limits the energy scale of bounce. Further unlike
that expected, the field during the contraction can not be driven
to arbitrary large value, even though the bounce occurs at Planck
scale.

\end{abstract}

 \maketitle

\section{Introduction}

The idea of oscillating universe is ancient, in which the universe
oscillates through many successive expansion/contraction cycles,
which was proposed originally by Tolman \cite{T} in the 1930's.
Recently this idea has been awaken again in cyclic scenario
\cite{KOS}, which is motivated by the string/M theory. The
relevant dynamics with primordial perturbations can be described
by an effective theory in which the separation of the branes in
the extra dimensions is modeled as a scalar field. But the
problems around bounce may remain. For contracting universe, it is
hard to escape from a singularity in the frame of general
relativity. Thus for the scenario of oscillating universe, one of
difficulties of implementing it is how to obtain a non singular
bounce. Several proposals have been discussed, such as using the
negative energy density \cite{PP, F, AWA} or curvature term
\cite{GT}, or some high order correction term in the action
\cite{AGN, GMV}.

During contraction of each cycles, the scalar field can be driven
and roll up along its potential to large enough value for
inflation to occur, which has been noticed in Ref. \cite{KSS} for
closed universe, in which it has been shown that an oscillating
universe certainly undergoes inflation after a finite number of
cycles. During this process, a contracting phase dominated by
kinetic energy followed by an inflation. Further, in Ref.
\cite{PFZ}, such scenario has been proposed to provide a possible
explanation for the observed low CMB anisotropies on large angular
scale. In this proposal, a kinetic dominated contracting phase is
matched to an inflation phase. The power spectrum during the
contraction $\sim k^3$ \footnote{This is also the usual results of
Pre Big Bang scenario \cite{GV}, see Ref. \cite{V} for a review.
}, which lead to an intense suppression of CMB quadrupole, in the
meantime the nearly scale-invariant spectrum from slow-roll
inflation is recover on small scale. But a physical mechanism for
bounce is not included. In this paper, we make use of possible
high energy correction to the Friedmann Eq. \cite{SS} to implement
the bounce. We firstly focus the behavior of massive scalar field
during such a bounce semianalytically and find that the slow-roll
inflation can be preceded by the kinetic dominated contraction and
the required e-folds number during the inflation after the bounce
limits the energy scale of bounce. Finally we check and confirm
these features numerically. Note that a similar scenario has been
studied in string inspired frame \cite{PTZ}, in which the
Gauss-Bonnet term is used to construct a non singular bounce and
the superinflation regime before slow-roll inflation leads to a
suppression of primordial spectrum. This has also a similarity to
the case in loop quantum gravity \cite{TSM}. The effects from loop
quantum gravity can driven the inflaton to its potential hill and
increase the parameter space of initial conditions for successful
inflation. Furthermore, the oscillatory universe induced by the
effects from loop quantum gravity was investigated in Ref.
\cite{ST}.

\section{Dynamics of Model}

In this section we study the cosmological behavior of massive
scalar field before and after a bounce phase semianalytically. We
use the modified Friedmann equation \be h^2 = {1\over 3}
(\rho_{\varphi} - {\rho_{\varphi}^2 \over \sigma}) \label{h2} \ee
to implement a realistic bounce, where $8\pi /m_p^2 =1$ has been
set, $\rho_{\varphi}$ is the energy density of scalar field and
$\sigma$ is the bounce scale, when $\rho_{\varphi}=\sigma$, the
universe bounces. Eq. (\ref{h2}) may be motivated in brane world
scenario \cite{SS}, see also \cite{BFK}, where our universe is
embedded in high dimension space/time, see Ref. \cite{M} for a
recent introduction. Further it has been pointed out in Ref.
\cite{CF} that due to the effects of the bulk and Israel matching
conditions there might be some forms $h(\rho)$ which is not
standard Fridmann-like.

The motion equation of scalar field $\varphi$ with the mass $m$ is
\be {\ddot \varphi} + 3h{\dot \varphi} +m^2\varphi =0
\label{phi}\ee
We assume that initially the universe is in contracting phase and
the field $\varphi$ is in the bottom of its potential. Thus
$\rho_{\varphi}$ is small and the term $\rho_{\varphi}^2$ in Eq.
(\ref{h2}) is negligible. Some possible fluctuations will make
$\varphi$ departure from its minimum and oscillate near
$\varphi=0$. Since in this case $m^2\gg h^2$, {\it i.e.} the
frequency of oscillation is much larger than the evolution rate of
universe, thus after implementing the change of variable $\varphi
=a^{-{3\over 2}} u$,  \be {\ddot u}+m^2 u \simeq 0 \ee can
approximately be obtained,
which can be solved as an oscillation \be u \simeq u_{a}
\sin{(mt)} \label{ophi}\ee where $u_{a}$ is a constant amplitude.
Thus we have \be \varphi= a^{-{3\over 2}} u_{a} \sin{(mt)}
\label{phimt}\ee When taking the time average over many
oscillations of field, $<p_{\varphi}>\simeq 0$ is obtained, which
is similar to the case dominated by matter. Thus in this regime
the universe contracts as $a \sim t^{2/3}$ and the energy density
of the field grows as \be \rho_{\varphi} \sim a^{-3} u_a^2
m^2\label{rho}\ee
From (\ref{rho}), we can see that with the contraction of universe
the energy density of scalar field increases. Thus when $h^2\sim
\rho_{\varphi}\gtrsim m^2$,
the term relevant with $h$ of Eq. (\ref{phi}) can no more be
neglected. In this case, $a^{-3} u_a^2 m^2 \gtrsim m^2$, which can
be reduced to $ \varphi_a^2\gtrsim a^{-3} u_a^2 \gtrsim 1$. Thus
as long as the oscillating amplitude of $\varphi$ field is larger
than $1$, the oscillation of field will end and the field will
roll up along its potential. Instead of Eq. (\ref{phimt}), \be
{\ddot \varphi}+3h{\dot \varphi}\simeq 0 . \label{kphi}\ee is
satisfied. From Eq. (\ref{kphi}), we have \be {\dot \varphi}
\simeq {c\over a^3} \label{phia}\ee where $c$ is the integral
constant. Thus with the further contraction the universe will
enter into the regime dominated by kinetic energy of field
rapidly, {\it i.e.} ${\dot \varphi}^2 \gg m^2\varphi^2$, where
$p_{\varphi}\simeq \rho_{\varphi}$. This leads to $\rho_{\varphi}
\sim {\dot \varphi}^2 \sim 1/a^6 $, thus we have \be a^3\simeq
c\sqrt{3\over 2}(t_{s}-t) \label{a3s}\ee where $t_{s}$ is the the
integral constant.
and \be \varphi \simeq \varphi_{k} + \sqrt{2\over
3}\ln{({t_{s}-t_{k}\over t_{s}- t})} \label{phibou}\ee where
$\varphi_{k}$ and $t_{k}$ are the value of field and time just
entering into the regime dominated by kinetic energy,
respectively.


When the kinetic energy of $\varphi$ approaches $\sigma$, we have
to includes the $\rho_{\varphi}^2$ correction in Eq. (\ref{h2}).
Thus instituting Eq. (\ref{phia}) into Eq. (\ref{h2}), we obtain
\be 6\left({{\dot a}\over a}\right)^2= {c^2\over
a^{12}}\left(a^6-{c^2\over 2\sigma}\right) \label{h4}\ee whose
solution can be obtained as follows \be  a^3 = \sqrt{{3 c^2\over
2} (t-t_{b})^2 + {c^2\over 2\sigma}}\label{a6}\ee where $t_{b}$ is
the integral constant determined by $\sigma$ as well as the value
of ${\dot \varphi}$ and $a$ at $t=t_{k}$. When $t=t_{b}$, the
scale factor of the universe arrives at its minimum and after
$t_{b}$ the universe bounces and expands.

For the universe dominated by stiff fluid with the same state
equation $p =\rho$ as the case dominated by kinetic energy of
scalar field, its evolution will symmetric before and after the
bounce, and up to all time. But the evolution of the universe
driven by massive scalar field is different from that with fluid.
After the bounce, the field $\varphi$ will still roll up. However,
since $h>0$, the term $3h{\dot\varphi}$ of Eq. (\ref{phi}) serves
as a damping term. Thus the roll-up motion of $\varphi$ will decay
quickly. When the velocity of $\varphi$ arrives at 0, it reverses
and rolls down along the potential. For the case of $\varphi
\gtrsim 1$, 
we have $m^2\lesssim h^2$. The field will enters the slow-roll
regime in which the universe is dominated by the potential energy
of the scalar field
and \be 3h{\dot \varphi} + m^2\varphi \simeq 0 ,\ee is satisfied.
In this case $p_{\varphi}\simeq -\rho_{\varphi}$ and the universe
inflates. The motion of field $\varphi$ is given by \be \varphi
\simeq \varphi_{r} -\sqrt{2\over 3} mt\ee where $\varphi_{r}$ is
the value of field at reverse. When $m^2\sim h^2$, {\it i.e.}
$\varphi\sim 1$, the inflation ends and the field $\varphi$ will
reenter into an oscillatory stage.

Therefore from above discussion, we can see that before the
bounce, the field initially in the bottom of its potential can be
driven by roll up its potential hill and after the bounce it
reverses and rolls down, and drives the inflation of universe.

\section{Numerical Results}

Then we study the above evolutive process numerically. We take $m
= 0.001$ and $\sigma = 0.0001$, and $a(0)=1$, $\varphi(0)=0$ and
${\dot \varphi}(0)=0.0001$ for initial values. The evolutions of
$a$ and $h$ are plotted in Fig. 1. We can see distinctly that the
bounce is implemented. In Fig. 2, we plot the evolution of field
$\varphi$. Initially the field oscillates about the bottom of its
potential many times, then the field $\varphi$ roll up along its
potential rapidly. The field still rolls up after the bounce, but
its velocity will approaches 0 quickly. Thus shortly it will
reverse and roll down, and then enter into slow-roll regime. When
the field decreases to $\sim 1$ linearly, the slow-roll ends and
the field will oscillates about $\varphi =0$ again.
These numerical results are consistent with our semianalytical
discussions in last section. The modification scale $\sigma$
determines the energy scale of bounce, thus in some sense also
determines the maximal value $\varphi_r$ which the $\varphi$ can
roll up to, see Fig. 3 for numerical solutions with various
$\sigma$. We can simply estimate the value of $\varphi_{r}$. When
the field just enter into the stage dominated by kinetic energy,
${\dot \varphi}_{k}^2\simeq m^2\varphi^2_{k}$ can be satisfied
approximately, and when the field is at the point of the bounce,
its value is ${\dot \varphi}_{b}^2 \simeq \sigma$. Thus from Eqs.
(\ref{phia}) and (\ref{a3s}), we have \be \left({t_{s}- t_{
b}\over t_{s}-t_{k}}\right)^2 = \left({a_{b}\over a_{ k}}\right)^6
= \left({{\dot \varphi}_{k}\over {\dot \varphi}_{
b}}\right)^2\simeq {m^2 \varphi_{k}^2\over \sigma}\ee Instituting
it into Eq. (\ref{phibou}), \be \varphi_{r} \lesssim\varphi_{k} +
\sqrt{2\over 3}\ln{({\sigma\over m^2
\varphi_{k}^2})}\label{phibou2}\ee can be obtained, where the
second term has been doubled since during kinetic dominated regime
the evolution of the field can be regarded approximately as
symmetrical before and after the bounce, which can also be seen
from Fig. 4. For a reasonable estimation, we take $\varphi_{
k}\sim 1$, thus obtain $\varphi_{r}\lesssim 4$ for $\sigma
=0.0001$ and $\varphi_{r}\lesssim 10$ for $\sigma =0.1$, which is
compatible with the numerical results of Fig. 3. In this case
there is a up-limit $\varphi_{r}\lesssim 11$ for $\sigma =1$, in
which the modification of Friedmann equation is at Planck scale.
Therefore, from Eq. (\ref{phibou2}), for massive scalar field
initially in the bottom of its potential with a small velocity
from fluctuations, the maximal value that it can roll up to during
the contraction is only dependent on its mass $m$ and the
modification scale $\sigma$. The larger $\varphi_{r}$ is, the
longer the time that the universe after the bounce inflates is.

\begin{figure}[]
\begin{center}
\includegraphics[width=8cm]{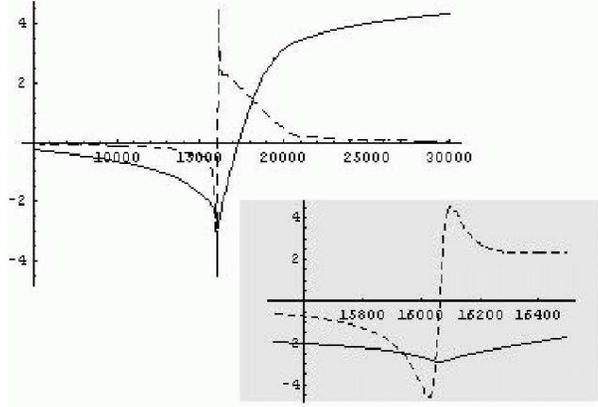}
\caption{ The scale factor $\ln{a}$ (solid line) and the Hubble
parameter $h$ (dashing line) as a function of time. The inset is a
zoom before and after the bounce. $h$ has been multiplied by 1000.
} \label{fig1}
\end{center}
\end{figure}

\begin{figure}[]
\begin{center}
\includegraphics[width=8cm]{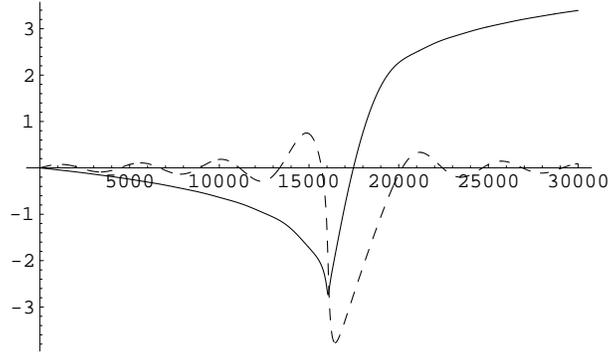}
\caption{ The scale factor $\ln{a}$ (solid line) and the field
$\varphi$ as a function of time.  } \label{fig1}
\end{center}
\end{figure}


\begin{figure}[]
\begin{center}
\includegraphics[width=8cm]{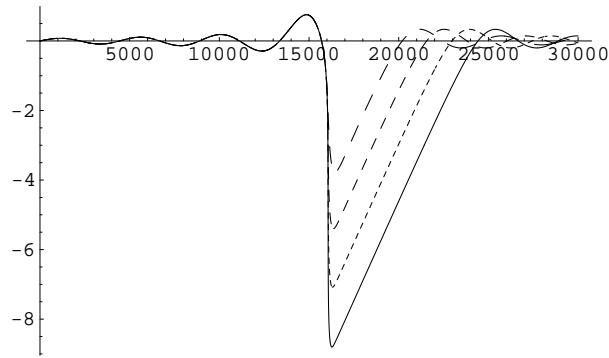}
\caption{ The field $\varphi$ as a function of time for various
$\sigma$. From long dashing line to solid line, $\sigma$ takes
0.0001, 0.001, 0.01, 0.1, respectively. Since $m$ is taken same,
initially the oscillation of fields has same frequency, which can
be seen from Eq. (\ref{ophi}). } \label{fig1}
\end{center}
\end{figure}

Focusing on the chaotic inflation model \cite{L1}, instead of
$8\pi/m_p^2 =1$ on whole paper, we take $m_p^2 =1$ here. The
e-folds number is given by \be N \simeq 2\pi \varphi_{r}^2 \ee and
Eq. (\ref{phibou2}) is modified as \be \varphi_{r}
\lesssim\varphi_{k} + \sqrt{1\over 12\pi}\ln{({\sigma\over m^2
\varphi_{k}^2})}\label{phibou3}\ee For $N\gtrsim 60$, this
requires $\varphi_{r} \gtrsim 3 $. Considering $m\sim 10^{-6}$
required by observational primordial perturbations, the limit to
$\sigma$ is given by $\sigma\gtrsim 10^{-8}$. Further for $\sigma
=1$ the maximal value that the field can roll up to is given by
$\varphi_{r}\lesssim 1+12\sqrt{1\over 12\pi} \ln{10} \simeq 5.5 $,
which is suitable for a successful inflationary cosmology but is
far away from eternal inflation \cite{L2}.

\begin{figure}[]
\begin{center}
\includegraphics[width=8cm]{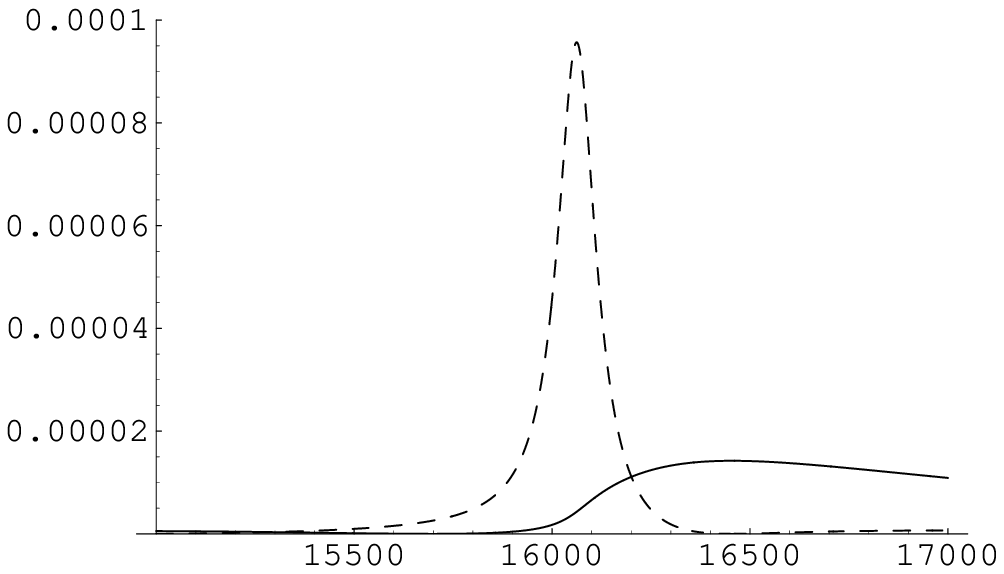}
\caption{The potential energy 
(solid
line) and the kinetic energy 
(dashing
line) of the field as a function of time before and after the
bounce.  } \label{fig1}
\end{center}
\end{figure}




\begin{figure}[]
\begin{center}
\includegraphics[width=8cm]{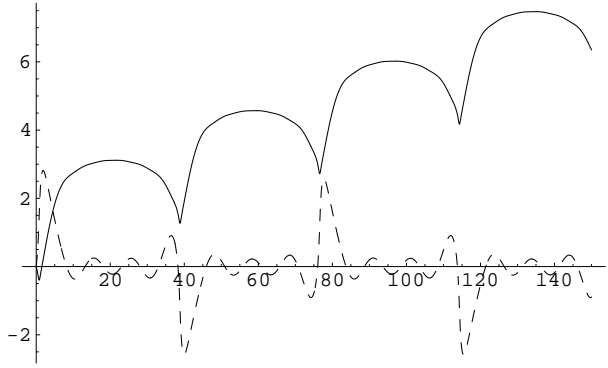}
\caption{The scale factor $\ln{a}$ (solid line) and the field
$\varphi$ (dashing line) as a function of time for an oscillating
universe. } \label{fig1}
\end{center}
\end{figure}

\begin{figure}[]
\begin{center}
\includegraphics[width=8cm]{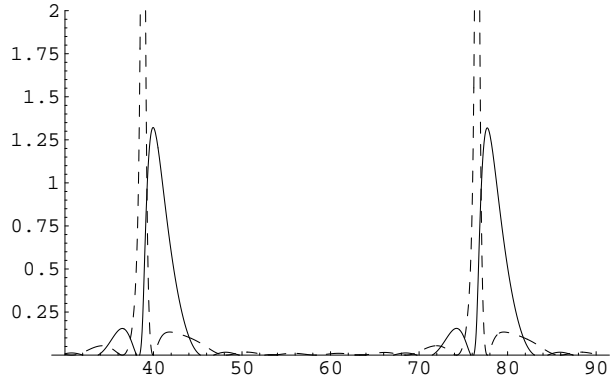}
\caption{ The potential energy
(solid line) and the kinetic energy 
(dashing line) of the field as a function of time for an
oscillating universe. The time taken only includes two cycles. }
\label{fig1}
\end{center}
\end{figure}

\begin{figure}[]
\begin{center}
\includegraphics[width=8cm]{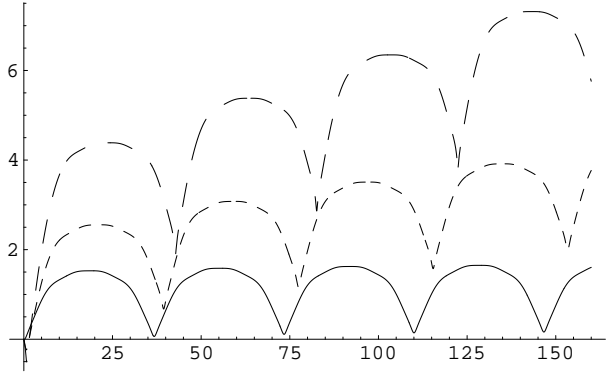}
\caption{ The scale factor $\ln{a}$ as a function of time for an
oscillating universe. The long, short dashing and solid lines are
the cases of ${\sigma\over m^2}= 50, 5, 2.5$, respectively. For
${\sigma\over m^2}=2.5$, from (\ref{phibou2}), we have
$\varphi_{r}\lesssim 1$. In this case, inflation dose not occur in
each cycle, thus the amplitude of successive cycle dose not
increase. } \label{fig1}
\end{center}
\end{figure}

We have found that the slow-roll inflation after the bounce can be
preceded by the kinetic dominated contraction. Further we would
like to check this feature in a controlled model of oscillating
universe \footnote{To realise a transition from expansion to
contraction, we take the density of potential energy $<0$ in the
minimum. The details of cosmology with negative potential have
been studied in Ref. \cite{FFKL}. }. We plot the evolutions of
$(a,\varphi)$ and the (kinetic, potential) energy with time in
Fig. 5 and Fig. 6, respectively. To show the oscillation of
universe better, we take $m^2 =0.2$, $\sigma =50$ and $a(0)=1$,
$\varphi(0)= 0$ and ${\dot \varphi} =1$ for initial values. We see
that the maximum of expansion amplitude in each successive
expansion/contraction cycle grows gradually. This asymmetry of
cosmological evolution seems conflicted with intuition, since both
the field equation (\ref{phi}) and Fridmann equation (\ref{h2})
are non dissipative and time reverse invariant. The reason of
asymmetry has been analyzed in Ref. \cite{KSS} for the case of
closed universe. However, in fact this is not dependent on the
curvature of universe, but the evolutive behavior of scalar field.
For $\varphi \gtrsim 1$, during the expansion, the state equation
$p_{\varphi} \simeq -\rho_{\varphi}$, thus the field is in
slow-roll regime and the universe expands exponentially {\it i.e.}
$a\sim e^{ht}$. However, during the contraction, the universe is
dominated by kinetic energy of scalar field. In this case, $a\sim
t^{1/3}$. Thus the time when the field rolls up to the value from
which it will roll down into slow-roll inflationary regime is more
shorter. It is the evolutive asymmetry of the field that leads to
the growth of successive expansion/contraction cycles. When that
$m^2$ is the same order as $\sigma$, from (\ref{phibou2}), we
obtain $\varphi_{r} \lesssim 1 $. In this case, the field can not
roll up to slow-roll regime. Thus the asymmetry between the
expansion driven by the potential of field and the contraction
driven by the kinetic term of field will not exist. We can see
from Fig. 7 that the amplitude of successive cycle does not
increase in this case.

\section{Conclusion}

We make use of possible high energy correction to the Friedmann
equation to implement a bounce. We study the behavior of massive
field and find that the slow-roll inflation can be preceded by the
kinetic dominated contraction. During this process, the field
initially in the bottom of its potential can be driven by the
anti-frictional force resulted from the contraction and roll up
its potential hill. We show that the required e-folds number
during the inflation after the bounce limits the energy scale of
bounce, for $N\gtrsim 60$, $\sigma \gtrsim 10^{-8}$ is required.
Further unlike that expected, in our case the field can not arrive
arbitrary large value, even though the bounce occurs at Planck
scale, for $\sigma =1$, we have $\varphi_{r} \simeq 5.5$. The
reason is that during the contraction the kinetic energy of the
field is much larger than its potential energy and the growth of
the field value is only logarithmical. When the kinetic energy
approaches $\sigma$, its potential energy is still far less, see
Fig. 3. These features have been checked and confirmed
numerically. Moreover, for possible oscillating universe model, we
point out that the successive increasing of expansion amplitude of
each cycle is actually depended on whether there exists inflation
in each cycle. For the oscillating universe without inflation in
its each cycle, its amplitude dose not increase with time.

\textbf{Acknowledgments} This work is supported in part by
K.C.Wang Postdoc Foundation and also in part by the National Basic
Research Program of China under Grant No. 2003CB716300.

\end{document}